\documentclass[referee,useAMS,usenatbib,usegraphicx]{mn2e}
\usepackage{color}

\title[Transient X-ray binary IGR J01583+6713]{Multiwavelength study of the transient X-ray binary IGR J01583+6713}
\author[Kaur et al.]{Ramanpreet Kaur $^{1}$\thanks{E-mail: raman@aries.ernet.in}, Biswajit Paul  $^{2}$, 
Brijesh Kumar $^{1,3}$, Ram Sagar $^{1}$ \\
$^1$ Aryabhatta Research Institute of observational sciencES (ARIES), Manora Peak, Naini Tal, 263\,129, India\\
$^2$ Raman Research Institute, C.V. Raman Avenue, Sadashivanagar, Bangalore 560\,080, India \\
$^3$ Departamento de Fi$\!^{'}\!$sica, Universidad de Concepcio$\!^{'}\!$n, Casilla 160-C, Concepcio$\!^{'}\!$n, Chile \\}
\begin{document}

\date{}

\pubyear{2007}

\maketitle

\label{firstpage}

\begin{abstract}
We have investigated multiband optical photometric variability and stability of the
H$\alpha$ line profile of the transient X-ray binary IGR J01583+6713.
We set an upper limit of 0.05 mag on photometric variations in the
{\it V} band over a time scale of 3 months. The H$\alpha$ line is found to 
consist of non-Gaussian profile and quite stable for a duration of 
2 months. We have identified the spectral type of the companion star to be B2 IVe while
distance to the source is estimated to be $\sim$ 4.0 kpc. Along with the
optical observations, we have also carried out analysis of X-ray data from
three short observations of the source, two with the {\it Swift}--XRT and one with the
{\it RXTE}--PCA. We have detected a variation in the absorption column density, from
a value of 22.0 $\times$ 10$^{22}$ cm$^{-2}$ immediately after the outburst down to 
2.6 $\times$ 10$^{22}$ cm$^{-2}$ four months afterwards. In the quiescent state, 
the X-ray absorption is consistent with the optical reddening measurement of
E(B - V) = 1.46 mag. From one of the {\it Swift} observations, during which the X-ray 
intensity was higher, we have a possible pulse detection with a period of 469.2 s. 
For a Be X-ray binary, this indicates an orbital period in the range of 216--561 
days for this binary system.
   
\end{abstract}

\begin{keywords}
stars: individual: IGR J01583+6713 -- binaries: general -- stars: pulsars: 
general -- stars: emission-line, Be -- X-rays: stars
\end{keywords}

\section{Introduction}

The ESA X--ray and $\gamma$--ray observatory {\it INTEGRAL} has discovered more than 
100 sources during its continuous monitoring of the sky, especially in the direction of 
the Galactic Center 
and its surroundings, since its launch in 2002. Many of the X-ray sources 
discovered by {\it INTEGRAL} are characterized by a hard X-ray spectrum, with little 
or no detectable emission in soft X-rays, since they are heavily absorbed 
by interposing material. X-ray characteristics of most of these sources 
indicate them to be High Mass X-ray Binaries (HMXBs) and in many cases this has 
been proved by the discovery of their optical counterparts (e.g. Masetti et al. 2006a, 
Negueruela et al. 2007). 

The X-ray transient IGR J01583+6713 was discovered by the IBIS/ISGRI imager on
board {\it INTEGRAL} during an observation of the Cas A region on Dec 6, 2005 
(Steiner et al. 2005). The source was detected with a mean flux of about
14 mCrab in the 20-40 keV band, while a null detection was reported in the higher energy
band of 40-80 keV. In subsequent {\it INTEGRAL} observations of the same field 
during Dec. 8-10, 2005, the X-ray flux was found to be decreasing on a 
timescale of days. {\it Swift} observations of the source
on December 13, 2005 identified IGR J01583+6713 to be a point object 
located at RA : 01$^\mathrm{h}$ 58$^\mathrm{m}$ 18$\fs$2
and DEC: +67$^{\circ}$ 13$^{\prime}$ 25${\farcs}$9 (J2000) with an uncertainty of 3.5 arc seconds
and its spectral analysis revealed that it is highly absorbed 
with $N_H$ approximately 10$^{23}$ cm$^{-2}$ as compared to the estimated 
galactic value of $N_H$ = 4.7 x 10$^{21}$ cm$^{-2}$ towards the same direction
(Kennea et al. 2005). Follow-up optical and IR observations identified the
X-ray source as a Be star with magnitudes - 
B=14.98, R=13.24 and I=12.12, displaying strong H$\alpha$ (EW 70 \AA) and 
weak H$\beta$ (EW 6 \AA) lines in emission and it was therefore proposed
as the optical/IR counterpart of the X-ray source (Halpern et al. 2005).
Recently, based on low resolution single epoch optical spectroscopy on December 23, 2006, 
Masetti et al. (2006b) classified the counterpart as an early spectral type
(O8 III or O9 V) Galactic ($\sim$ 6.4 kpc) star and ruled out both the possibility of a 
supergiant companion and of the source being an LMXBs or CVs. 

In this paper, we present results of new optical (photometric and spectroscopic) 
observation of the Be system over an extended period and analyze all the X-ray data 
on IGR J01583+6713 present in the HEASARC archive. Section 2 and 3 present the
observations and data reduction. Section 4 deals with the results and discussions on 
the nature of the companion star, possible variability and characterization of 
the X-ray spectrum. We summarize the results in the last section.  

\section{Optical Observations}

The broadband photometric and intermediate resolution spectroscopic optical 
observations are described in the following sections.

\subsection{{\it UBVRI} Photometry}

We have carried out broadband Johnson {\it UBV} and Cousins {\it RI} CCD photometric 
follow-up observations on 13 nights from December 13, 2005 to March 06, 2006
using a 2k $\times$ 2k CCD camera mounted at f/13 Cassegrain focus of the $104$-cm
Sampurnanand Telescope (ST) at ARIES, Nainital, India. At the focus, a 24
$\mu$m square pixel of the 2048$\times$2048 size CCD chip corresponds to
$\sim$ 0.36 arcsec and the entire chip covers a square area of side
$\sim$ 13.0 arcmin on the sky. The read-out noise of the system is 5.3 electrons
with a gain factor of 10 electrons per analog-to-digital unit (ADU). During
our observations the seeing varied from about 1.2 to 2.4 arcsec. Table 1
lists the log of optical photometric observations of IGR J01583+6713 in 
{\it UBVRI} wavebands along with the number of frames taken and exposure time 
in the respective filters. Usually more than two exposures were taken in each filter
with a typical exposure times of 300 s, 300 s, 300 s, 200 s and 150 s in the {\it UBVRI} 
wave-bands respectively. Bias frames were taken intermittently and flat-field 
exposures were made of the twilight sky. In addition, as part of other 
ongoing programs, we could secure observations of Landolt (1992) standard fields 
in {\it UBVRI} wavebands on four nights (see Table 1) near zenith and the same field 
was also observed at about five different zenith distances for extinction 
measurements. 

Figure 1 shows the field chart of IGR J01583+6713 observed with ST on December 13, 2006
in the {\it B} waveband. Although all the observations were taken with the entire CCD chip 
of 13 $\times$ 13 arcmin, we have shown only a 7.5 $\times$ 7.5 arcmin CCD frame 
in Figure 1 for a closer look at the field. For photometric comparison we selected 
six stars with similar brightness and these are marked as 1, 2, 3, 4, 5 and 6 in 
Figure 1 while `T' denotes the X-ray transient IGR J01583+6713. 

Photometric data reductions were performed using the standard routines in 
{\small IRAF\footnote[1]{{\small IRAF} is distributed by National Optical Astronomy 
Observatories, USA}} and {\small DAOPHOT} (Stetson 1987). The zero-points and color 
terms were applied in the usual manner (Stetson 1992).
The standard magnitudes of all six stars were found to be stable on all four 
nights and were treated as local standards. Their mean standard magnitudes 
are listed in Table 2. The mean magnitudes and colors for IGR J01583+6713 are
estimated as $V=14.43\pm0.03$, $U-B=0.09\pm0.04$, $B-V=1.22\pm0.07$, $V-R=0.99\pm0.03$, 
$V-I=1.88\pm0.05$. The difference in the measured (B$_{obs}$, V$_{obs}$, R$_{obs}$ and  
I$_{obs}$) and standard (B$_{st}$, V$_{st}$, R$_{st}$ and I$_{st}$) {\it BVRI} magnitudes 
of IGR J01583+6713 and comparison star 1 are plotted in Figure 2. The variation was 
observed to be consistent with the typical photometric uncertainty with standard 
deviations of 0.01 mag for {\it BVR} and 0.02 mag for the {\it I} passband for comparison star 1. 
For IGR J01583+6713 an increase in flux of about 0.05 mag is apparent around 13 Dec 2005, followed 
by a constant flux level thereafter. 

\subsection{Spectroscopy}

Spectroscopic observations of IGR J01583+6713 were made on eight nights
during August 18, 2006 to October 28, 2006 using the Himalayan Faint Object
Spectrograph and Camera (HFOSC) available with the 2-m Himalayan Chandra
Telescope (HCT), located at Hanle, India. The pixel size of the CCD used is
15$\mu$m with an image scale of 0.297 arcsec pixel$^{-1}$. The log of 
observations with the respective exposure times is given in Table 3. We have
used a slit of dimensions 1.92$^{''}$x 11$^{'}$ for the H$\alpha$ line profile
study and a slit of dimensions 15.41$^{''}$x 11$^{'}$ for the calibration
observations. The spectrum were obtained with two different gratings (Grism 7 \& 
Grism 8) in the wavelength range of 3500-7000 \AA~and 5800-8350 \AA~at a 
spectral dispersion 
of 1.45 \AA~pixel$^{-1}$ and 1.25 \AA~pixel$^{-1}$ respectively. Data were
reduced using the standard routines within {\small IRAF}. The data were bias-corrected,
flat-fielded and the one dimensional spectrum extracted using the optimal
extraction method (Horne 1986). The wavelength calibration was done using FeAr and FeNe
lamp spectrum for the Grism 7 and Grism 8 spectrum respectively. We employed the
{\small IRAF} task {\it identify} and typically around 18 emission lines of Fe, Ar and Ne
were used to find a dispersion solution. A fifth order fit was used to achieve
a typical RMS uncertainty of about 0.1 \AA. For flux calibration 
of the IGR J01583+6713 spectrum in both Grism 7 and Grism 8, the instrumental response curves 
were obtained using spectrophotometric standards (Hiltner 600, Feige 110) 
observed on the same night, and the star's spectrum were brought to the relative flux scale.
H$\alpha$ is detected with a FWHM of $\sim$ 11.3 \AA~for all the spectrum
taken with Grism 8 of HCT. 

The combined flux calibrated spectrum of X-ray binary IGR J01583+6713,
taken with Grism 7 and Grism 8 on October 15, 2006 is shown in Figure 3 
over a  wavelength range of 3800--9000~\AA. The identified spectral 
features are marked in the spectrum. 
The blue region of Grism 7, in the wavelength region 3600-3800 \AA~
has poor Signal-to-Noise ratio and is not shown in Figure 3. 
Along with strong H$\alpha$ and H$\beta$, we also detected a few 
weak spectral features, mainly singly ionized Iron, neutral Helium 
and neutral Oxygen lines, in the IGR J01583+6713 continuum spectrum 
in the wavelength region 6100--7900 \AA, shown in Figure 4. 

\section{X-ray Observations}

The X-ray observations of IGR J01583+6713 were carried out with the X-ray Telescope (XRT) 
onboard the {\it Swift} satellite and with the Proportional Counter Array (PCA) onboard
the {\it RXTE} satellite. 

{\it Swift} carried out observations of 
IGR J01583+6713 on December 13, 2005 for 47 ks and on April 11, 2006 for 37 ks. 
Both the XRT observations had useful exposures of $\sim$ 8 ks. The standard data pipeline
package (XRTPIPELINE v. 0.10.3) was used to produce screened event files.
Only data acquired in the Photon counting (PC) mode were analyzed adopting
the standard grade filtering (0-12 for PC) according to XRT nomenclature.
X-ray events from within a circular region of radius 0.8 arcmin, centered at
the X-ray transient, were extracted for timing and
spectral analysis. Background data were extracted from a neighbouring
source free circular region of the same radius as taken for the source. Source and background
lightcurves were generated using the X-ray counts from the respective circular
regions with the instrumental time resolution of 2.5074 s. A final source
lightcurve was produced by subtracting the background lightcurve. The observations
were made for small segments and no variability beyond statistical variation is seen
in the X-ray light curves between the segments. 
The final source spectrum were obtained by subtracting the background spectrum,
and spectral analysis was done using the energy response of the
detector for the same day. 
 
{ \it RXTE} pointed observations were performed on December 14, 2005 with a total 
effective exposure time of $\sim$ 3 ks. The { \it RXTE}--PCA standard 2 data with a
time resolution of 16 s were used to extract the spectrum of the source in the 
energy range 3--20 keV. During
this observation only two PCUs were operational. The background spectrum for this
observation was generated using the task `pcabackest'. PCA background models
for faint sources were used to generate the background spectrum. The spectral
response matrix of the detector was created using the task `pcarsp' and
applied for spectral fitting. The spectrum was rebinned to have sufficient
signal to noise ratio in each bin of the spectrum. The source was also 
regularly monitored by the All Sky Monitor (ASM) on board { \it RXTE}. The ASM lightcurve 
is shown in Figure 5 from MJD 53660 to MJD 54160, including the outburst 
detected on MJD 53710. The epoch of detection of the hard X-ray transient 
IGR J01583+6713 is marked as `T' in the Figure 5. The soft X-ray enhancement 
in the {\it RXTE}--ASM light curve is consistent with a 10--20 mCrab intensity. 
The optical and X-ray observations of IGR J01583+6713 taken for the 
present study are also marked in Figure 5. 
 
We have used XSPEC12 for X-ray spectral analysis. We fitted both
a powerlaw model and a black--body model with line of sight absorption to the 
source spectrum obtained from { \it Swift} and { \it RXTE}. The spectral parameters 
obtained for both the models for the three observations are given in 
Table 4. Both the  powerlaw and black--body models fit the data well with 
a reduced $\chi^2$ in the range of 0.8--1.4, except for the { \it RXTE} spectrum 
which is not well fitted by a black--body 
model. Models are indistinguishable from spectral fit only. 
Figure 6 shows the powerlaw fit to the { \it Swift} observations made on December 13, 
2005 (top panel), { \it RXTE} observations made on December 14, 2005 (middle panel), 
and {\it Swift} data taken on April 05, 2006 (bottom panel). 

The {\it Swift} spectrum for both the observations made on December 13, 2005 and 
April 05, 2006 have coarse energy binning, making the Iron emission line at 6.4 keV 
undetectable in the raw spectrum. To find the upper limit on the 6.4 keV emission line 
in the {\it Swift} spectrum taken on December 13, 2005, we fixed the line-center 
energy at 6.4 keV in the spectrum and fitted a Gaussian to the line. The upper limit 
on the equivalent width of the 6.4 keV emission line is determined to be 100 eV 
with 90\% confidence limit. 

We searched for X-ray Pulse periods using a pulse-folding technique. The background 
count rate was subtracted from the IGR J01583+6713 lightcurve and the barycenter 
correction was done. The time resolution of the instrument and the time span of the continuous 
data constrained the pulse period search to the range 5s - 800s. We found X-ray 
pulsations with a pulse period of 469.2 s and with a pulsed fraction of 22\%. Figure 
7 shows the light curve of IGR J01583+6713, observed with {\it Swift} on December 
13, 2005, folded with the pulse period of 469.2 s. However the evidence of pulsation
detection is marginal with a false-alarm-probability of 10$^{-4}$.

\section{Results and Discussions}

\subsection{Photometric Variability}

There were 7 photometric observations made in 15 days from December
13, 2005 to December 28, 2005 and later on there were observations with a gap
of 20--25 days for the next 3 months. No variability of more than 2 sigma is found in any
pass-band but a larger variation and decreasing trend is seen for the first few
days of observations in all of the pass-bands, as shown in Figure 2. The scatter 
in data points of IGR
J01583+6713 is more than the comparisons stars by $\sim$ 2 sigma in all the pass-bands
and some increase in magnitude of about 0.05 mag can be seen around MJD 53725.
The decrease in reprocessed optical emission for few days of outburst does
indicate that source flux was decreasing during that period. However, we
claim no strong variability in this source and we cannot ascertain whether
the source was brighter in the optical band during the peak of its X-ray
outburst.  

\subsection{Stability of the H$\alpha$ emission line}

The dynamical evolution of Be envelope can be studied with the help of changes
in emission line profile (Negueruela et al. 2001). Table 3 shows the
spectroscopic observations of IGR J01583+6713 made over a period of 2 months.
Strong H$\alpha$ and H$\beta$ emission line features are always found in the
spectrum with equivalent width of -74.5 $\pm$ 1.6 \AA~and -5.6 $\pm$ 0.3 \AA~
respectively. This ratio is somewhat different from that reported by
Masetti et al (2006b), which may indicate structural changes in the circumstellar 
disk of Be star in this system.  

The line equivalent width and line profile of the H$\alpha$ line are found 
to be constant for the observations, listed in Table 3. The resolution of 
our instrument was not good enough to carry out a detailed study of the H$\alpha$ 
line shape. The H$\alpha$ line gives a poor fit with a Gaussian model, leaving the
wings of the line unfitted. However, the H$\alpha$ line, being very strong, of 
equivalent width -75 \AA, confirms that it originated in the circumstellar disk.
 
\subsection{X-ray variability and Spectrum}

X-ray observations of IGR J01583+6713 were made on December 13, 2005 by {\it Swift}
after the detection of its outburst on December 6, 2005 by {\it INTEGRAL}. 
The pulsations at 469.2 seconds were detected with a pulsed fraction of 22\%. 
Corresponding to 469.2 sec pulse period, we have estimated the X-ray binary
IGR J01583+6713 orbital period in the range 216--561 days (Corbet 1986) assuming the
maximum eccentricity of the orbit to be 0.47 for Be binaries as observed by Bildsten et al. (1997).

Figure 6 (top) and (bottom) show the powerlaw fitting to the {\it Swift}
observations made on December 13, 2005 and April 05, 2006. N$_H$ is found to 
decrease from 22.0 $\times$ 10$^{22}$ cm$^{-2}$ to 2.6 $\times$ 10$^{22}$ 
cm$^{-2}$ for the two {\it Swift} observations of IGR J01583+6713, using the powerlaw 
model. Using the blackbody model, N$_H$ is found to decrease from 
15.2 $\times$ 10$^{22}$ cm$^{-2}$ to 0.5 $\times$ 10$^{22}$ cm$^{-2}$. Photon index
and black-body temperature are in the range 1.7--2.0 and 1.3--1.8 keV respectively,
for all the observations listed in Table 4. {\it Swift} observations
clearly show that N$_H$ has decreased by about an order of magnitude in a span
of four months. The observed flux in 2-10 keV band has also decreased by a factor of $\sim$ 4.
Within measurement uncertainty, the power-law photon index (or the black--body
temperature in the blackbody emission model) is found to be unchanged.

Lack of variability in the optical photometric measurements indicate an
absence of changes in the distribution of circumstellar material around 
the companion star. However, from the X-ray spectral measurements we have 
detected a significant change in the absorption column density. During the first 
{\it Swift} observations, when the source was brighter, we measured an 
absorption column density of (22 $\pm$ 10) $\times$ 10$^{22}$ cm$^{-2}$ and 
the spectrum shows a upper limit of 100 eV on equivalent width of an iron K-fluorescence line.
If we assume an isotropic distribution of absorbing matter around the compact 
object, where the X-ray emission originates, we expect the Fe 6.4 keV spectral 
line to be detected with an equivalent width of 250 eV (Makishima et al. 1986 ), 
a factor of 2.5 more than the upper limit. The absence of a strong iron emission 
line indicates that the X-ray absorbing material around the compact object has 
a non-isotropic distribution, probably related to its disk structure.

\subsection{Spectral Classification}

A low dispersion (4\AA/pix) optical (3500-8700\AA) spectrum taken one day 
after the outburst was presented by Massetti et al. (2006b), However, due 
to the absence of any photospheric absorption feature and poor signal, they 
could not secure a definite classification so assuming a Galactic 
reddening and $(B-R)_{0}$ intrinsic color they assigned a spectral type 
of O8III or O9V. The present spectrum cover the same spectral region and 
have a better spectral resolution ($\sim$ 3 \AA~near H$_{\alpha}$), however, 
the MK classical region ($<$ 5000\AA) was too weak to identify absorption 
features. Figure 3 shows the combined flux calibrated spectrum of 
IGR J01583+6713 on October 15, 2006. 
The continuum normalized spectrum taken on October 15, 2006 and 
October 16, 2006 are shown in Figure 4. Most of the spectral features 
in the 3800--8800 \AA~region are identified and marked. We compare 
these features with a near infrared spectral library by Andrillat 
(1988 \& 1990, henceforth AND90) of a sample of 70 emission line Be stars 
with known MK spectral type (B0--B9) at resolution (1.2 \AA). We describe 
the spectrum as following.

Hydrogen lines are seen in emission. H$_{\alpha}$ and H$_{\beta}$ show single 
peak while the Pashcen lines (P12--P20) have a double-peak structure with the red 
(R) peak greater than the blue (V) one. CaII (8498, 8542, 8662~\AA) and NI 
(Multiplet at 8630 \AA~and 8680\AA~) are also in emission. In their sample, 
AND90 found that the HI and CaII emission features are strong in early type 
($<$ B5) stars and diminish strongly for later type stars. CaII triplet 
emission was associated with a large IRAS excess.

The present spectrum also show OI (8446\AA~and Triplet 7773\AA~) in emission
which is seen in most of the cases with stars having spectral type earlier 
than B2.5. The features of IGR J1583+6713 resemble most closely to HD 164284 
(B2 IV-Ve) in Andrillat (1988) and HD 41335 (B2 IVe) in Andrillat (1990).
For a comparison, HD 164284 spectrum is shown on the top of IGR J01583+6713 
spectrum in Figure 8 over a wavelength range of 7500--8800\AA.
 
FeII (6248, 6319, 6384, 6456, 6516, 7515 and 7711 \AA) and SiII (6347 \AA) lines
appear in emission. He I lines at 6678 and 7281 are in absorption with
asymmetric profiles indicating the presence of an emission region. Based on 
these characteristics it is suggested that its spectral type lies around B1 to B3,
however a later spectral type cannot be completely ruled out and a further
high resolution spectrum in the Optical and NIR would be required to ascertain the
true spectral type for the transient. Furthermore, the weak Fe 6.4 keV line (EW $\le$ 100 eV) 
in the X-ray spectrum and a high hydrogen column density may suggest the presence of 
a wind-powered accretion disc, hence leaving the possibility of a 
low-luminosity blue supergiant open. However, the strong Hydrogen lines suggest 
that it belongs to a main sequence (IV-V) luminosity class (Leitherer 1988).   

We employed the reddening free Q-parameter to further ascertain the spectral 
type of the star. Using the normal reddening slope, X[E(U-B)/E(B-V)]=0.72, the 
Q[=(U-B)-X(B-V)] parameter was found to be -0.63 $\pm$ 0.06 and this corresponds
to a spectral type of B2-3 for an early type main-sequence star (Johnson \& Morgan
1954). 

We derive a color excess E(B - V) = 1.46 $\pm$ 0.05 mag by adopting B2 IV as a companion and 
taking the intrinsic color (B - V)$_0$ = -0.24 mag (Schmidt-Kaler 1982) and a mean observed 
color of 1.22 $\pm$ 0.05 mag. An estimate of reddening at other wavelengths is made
following Fitzpatrick et al. (1999) and the intrinsic colors are found to be
(U - B)$_0$ = -1.00, (V - R)$_0$ = -0.15 and (V - I)$_0$ = -0.35 quite consistent 
with intrinsic colors corresponding to a B2 IV star with (U - B)$_0$ = -0.86 (Schmidt-Kaler 1982),
(V - R)$_0$ = -0.10 and (V - I)$_0$ = -0.29 (Wegner 1994). The above color excesses 
yield a value of visual extinction $A_{V}$ = $4.5 \pm 0.2$ mag. Using the absolute magnitude 
from Lang et al. (1992), and from the relation m - M = 5logD - 5 + A$_V$, the distance to 
the source is estimated to be 4.0 $\pm$ 0.4 kpc placing the transient well beyond the Perseus 
arm of the Milky Way.

For A$_V$ $\sim$ 4.5, the corresponding column density N$_H$ is $\sim$ 8.1 $\times$ 
10$^{22}$ cm$^{-2}$. The X-ray observations made by {\it Swift} during the quiescent state of the
transient IGR J01583+6713 on April 05, 2006 gave N$_H$ of the order of 
0.5 $\times$ 10$^{22}$ cm$^{-2}$ for blackbody fit and 3 $\times$ 10$^{22}$ cm$^{-2}$
for the powerlaw fit. The galactic HI column density is 0.4 $\times$ 10$^{22}$ 
cm$^{-2}$ in the direction of the transient. However, the N$_H$ obtained using the powerlaw model is 
closer to the N$_H$ calculated using optical extinction measurement as compared to
the blackbody model. Thus we conclude that the powerlaw model fits better to the 
X-ray data than the blackbody spectrum. The powerlaw index is found to be varying between 
1.7--2.0, which is common in accretion powered X-ray pulsars. 

\section{Summary}

The nature of the hard X-ray transient IGR J01583+6713 and its optical counterpart
has been investigated using new photometric and spectroscopic data in the optical as well as
the X-ray band. Over a one year period since the X-ray outburst (December 06, 2005), UBVRI 
CCD photometric data were collected on 15 nights using 1-m ST optical telescope at
Nainital. The intermediate resolution (3\AA~ at H$\alpha$) spectrum in the 
3800--8800\AA~region was monitored with the 2-m HCT telescope at Hanle, India 
on 8 nights. In X-ray, the archival spectral and timing data from {\it Swift} and 
{\it RXTE} observations were analysed to probe the nature of this 
highly obscured X-ray source. The main conclusions of the study are as follows. 

\begin{enumerate}

\item
No significant variability in V-band is seen, however an upper
limit of 0.05 mag is set over a time scale of 3 months since the X-ray outburst.

\item
The spectral characteristics of the optical counterpart were found to be
consistent with a B2 IV (classical Be) star showing strong emission lines of Hydrogen 
(single peak Balmer and double peak Paschen), ionized Calcium, 
ionized Silicon, Oxygen, Nitrogen and ionized Iron. The source is
located (l=129$^{\circ}$, b = 5$^{\circ}$) well beyond the Perseus arm of the Milky Way at a 
heliocentric distance of $\sim$ 4.0 $\pm$ 0.4 kpc.   

\item
We derive a hydrogen column density of 22.0 $\times$ 10$^{22}$ cm$^{-2}$ from 
the X-ray spectrum taken with {\it Swift} just after the outburst. The quiescent 
phase column density was found to be consistent with the optical extinction 
measurements. The timing measurement suggest a pulse period of 469.2 second and 
an orbital period in the range of 216--561 days for this binary system. 

\end{enumerate}

\section*{Acknowledgments}
We thank the referee "Juan Fabregat" for suggestions that helped us to improve 
the paper. We thank Dr. Maheswar Gopinathan and Mr. Manash Samal for the useful 
discussions on spectroscopy. One of the authors (Ramanpreet) thanks Jessy 
Jose, and Kuntal Misra for their help with the photometric data reduction
and Dr. Anna Watts for kindly going through the draft version of the paper.
This research has made use of data obtained through the High Energy 
Astrophysics Science Archive Research Center Online Service, provided 
by the NASA/Goddard Space Flight Center.

%Tables

%Table 1
\clearpage
\begin{table*}
\caption{Log of broadband optical photometric observations of the transient source and 
Landolt (1992) standard fields}
%\medskip
\scriptsize
\begin{tabular}{llll} \hline
Object &Date of	   &   Filter    &  Exposure time              \\
& Observation&           & (seconds)            \\ \hline
IGR J01583+6713 &&& \\
&2005 Dec 13&  V/R        & 2$\times$600/2$\times$300      \\     
&2005 Dec 17&  B/V/R/I    & 2$\times$300/2$\times$300/3$\times$200/2$\times$150\\ 	
&2005 Dec 18&  B/V/R/I    & 3$\times$300/3$\times$300/3$\times$150/3$\times$150\\  	
&2005 Dec 19&  V/R/I      & 1$\times$300/1$\times$200/1$\times$150\\
&2005 Dec 20&  B/V/R/I    & 1$\times$300/2$\times$300/1$\times$300/3$\times$150\\
&2005 Dec 21&  R/I        & 1$\times$150/2$\times$150      \\
&2005 Dec 28&  B/V/R/I    & 1$\times$300/2$\times$300/2$\times$200/3$\times$150\\
&2006 Jan 25&  B/R/I      & 1$\times$300/2$\times$200/1$\times$150      \\
&2006 Jan 26&  B/V/R/I    & 1$\times$300/2$\times$300/2$\times$200/2$\times$150\\
&2006 Feb 24&  B/V/R      & 2$\times$300/2$\times$300/2$\times$200      \\
&2006 Feb 28&  V/R        & 1$\times$300/1$\times$200      \\
&2006 Mar 02&  U/B/V/R/I  & 2$\times$300/2$\times$300/2$\times$300/2$\times$200/2$\times$150\\
&2006 Mar 06&  B/V/R      & 1$\times$300/1$\times$300/1$\times$200      \\
&2006 Nov 24&  U/B/V/R/I  & 2$\times$300/2$\times$300/2$\times$300/2$\times$200/2$\times$150\\
&2006 Dec 13&  U/B/V/R/I  & 3$\times$300/3$\times$300/3$\times$300/2$\times$200/2$\times$150\\
Landolt standard field  &&& \\
SA104 & 2006 Jan 25 &B/R/I & 11$\times$300/11$\times$60/11$\times$60\\
SA101 & 2006 Mar 02 &U/B/V/R/I & 9$\times$300/9$\times$180/9$\times$180/9$\times$120/9$\times$120\\
SA92 & 2006 Nov 24 &U/B/V/R/I & 2$\times$300/2$\times$300/2$\times$180/2$\times$130/2$\times$100\\
SA95 & 2006 Nov 24 &U/B/V/R/I & 7$\times$500/7$\times$300/7$\times$150/7$\times$100/7$\times$100\\
SA98& 2006 Dec 13 &U/B/V/R/I & 7$\times$450/7$\times$300/7$\times$120/7$\times$60/7$\times$60\\ 
RU149& 2006 Dec 13 &U/B/V/R/I & 2$\times$300/2$\times$300/2$\times$120/2$\times$60/2$\times$60\\ \hline

\end{tabular}      
\end{table*}

\begin{table*}
\caption{{\it BVRI} magnitudes of comparison stars}
%\medskip
\scriptsize
\begin{tabular}{lllll} \hline
Star no.     &         B   &    V       &     R      &    I \\ \hline
1            &     15.55  &  14.51    &   13.91   &   13.37 \\ 
2            &     17.57  &  15.45    &   14.17   &   13.00 \\
3            &     16.54  &  15.38    &   14.68   &   14.02 \\
4            &     16.25  &  14.92    &   14.11   &   13.36 \\
5            &     16.13  &  14.73    &   13.90   &   13.20 \\
6            &     17.71  &  15.52    &   14.20   &   12.99 \\ \hline
\end{tabular}      
\end{table*}     
 
\clearpage

\begin{table}
\begin{minipage}{110mm}
\caption{Log of spectroscopic observations of the X-ray transient IGR J01583+6713}
\scriptsize
\begin{tabular}{ccc} \hline
Date of    &  Wavelength Range              &  Exposure time              \\
Observation& (\AA{})             &  (seconds)            \\ \hline
2006 Aug 18&  3500-7000:5200-9200&  1x900/1x900\\
2006 Oct 14&  3500-7000:5200-9200&  1x900/1x900\\
2006 Oct 15&  3500-7000:5200-9200&  1x1200/1x900\\
2006 Oct 16&  5200-9200          &  1x900\\
2006 Oct 17&  5200-9200          &  1x900\\
2006 Oct 18&  3500-7000:5200-9200&  1x900/1x900\\
2006 Oct 28&  5200-9200          &  1x900\\
2006 Oct 29&  5200-9200          &  1x600\\
\hline
\end{tabular} 
\end{minipage}
\end{table}

\begin{table}
\begin{minipage}{110mm}
\caption{Spectral Parameters for {\it Swift} and { \it RXTE} observations. } 
\scriptsize

\begin{tabular}{lccc} \hline
& \multicolumn{1}{c}{{ \it RXTE}}                              & \multicolumn{2}{c}{{\it Swift}} \\ \hline
& \multicolumn{3}{c}{Spectral parameters of model (Absorption + powerlaw)}    \\ \hline
Parameter                                          & 2005 December 14          & 2005 December 13     & 2006 April 05              \\ \hline 
N$_H$ x 10$^{22}$ (cm$^{-2}$)                      & $<$ 6.0 (with 90\% confidence)& 21.96 $\pm$ 9.60     & 2.58 $\pm$ 1.06            \\ 
PhoIndex                                           & 1.71 $\pm$ 0.22           &  1.74 $\pm$ 0.56     & 2.05 $\pm$ 0.60            \\ 
Reduced $\chi^2$/dof                               & 1.4/14                    &  0.8/19              & 1.1/13                     \\ 
Observed Flux  2-10 keV (ergs cm$^{-2}$ s$^{-1}$)  & 5.0e-12                   & 11.6e-12             & 3.1e-12                   \\ 
Unabsorbed Flux 2-10 keV  (ergs cm$^{-2}$ s$^{-1}$)& 5.5e-12                   & 29.7e-12             & 4.0e-12                   \\ \hline
& \multicolumn{3}{c}{Spectral parameters of model (Absorption + blackbody)}                                                        \\ \hline
Parameter                                          & 2005 December 14          & 2005 December 13     & 2006 April 05              \\ \hline
N$_H$ x 10$^{22}$ (cm$^{-2}$)                      & $<$ 1.5 (with 90\% confidence)& 15.16 $\pm$ 6.23     & 0.45    $\pm$ 0.22         \\ 
Blackbody temperature (keV)                        & 1.61 $\pm$ 0.44           &  1.81 $\pm$ 0.49     & 1.34    $\pm$ 0.10         \\ 
Reduced $\chi^2$/dof                               & 2.5/14                    &  0.8/19              & 1.2/13                     \\ 
Observed Flux 2-10keV (ergs cm$^{-2}$ s$^{-1}$)    & 5.1E-12                   & 10.8E-12             & 3.7E-12                    \\ 
Unabsorbed Flux 2-10keV  (ergs cm$^{-2}$ s$^{-1}$) & 5.8E-12                   & 19.6E-12             & 3.9E-12                    \\ \hline

\end{tabular} 
\end{minipage}
\end{table}

\clearpage

\begin{figure}
\centering
\includegraphics[scale=0.7]{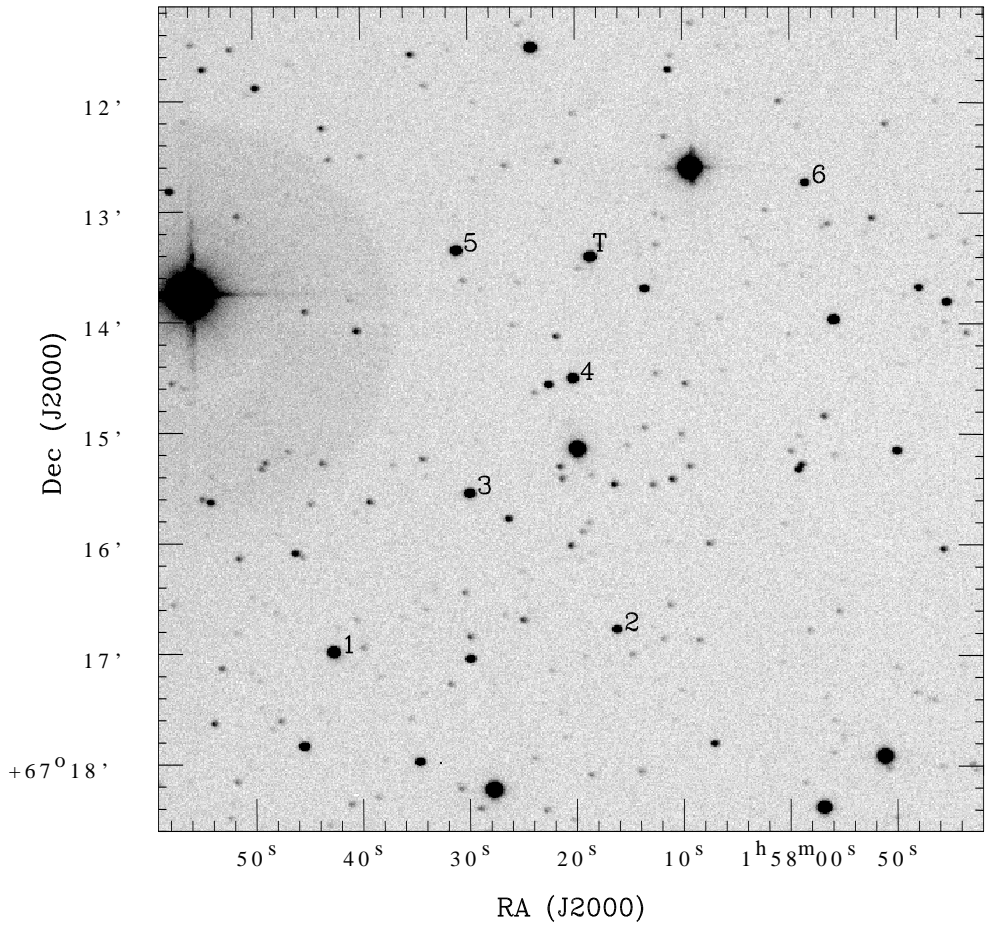}
\caption{Identification chart of IGR J01583+6713 taken with ST in B passband on December 13, 2006. The transient is marked
as `T' and comparison stars are marked as 1, 2, 3, 4, 5 and 6 in the figure}
\end{figure}

\clearpage

\begin{figure}
\centering
\includegraphics[height=2.8in]{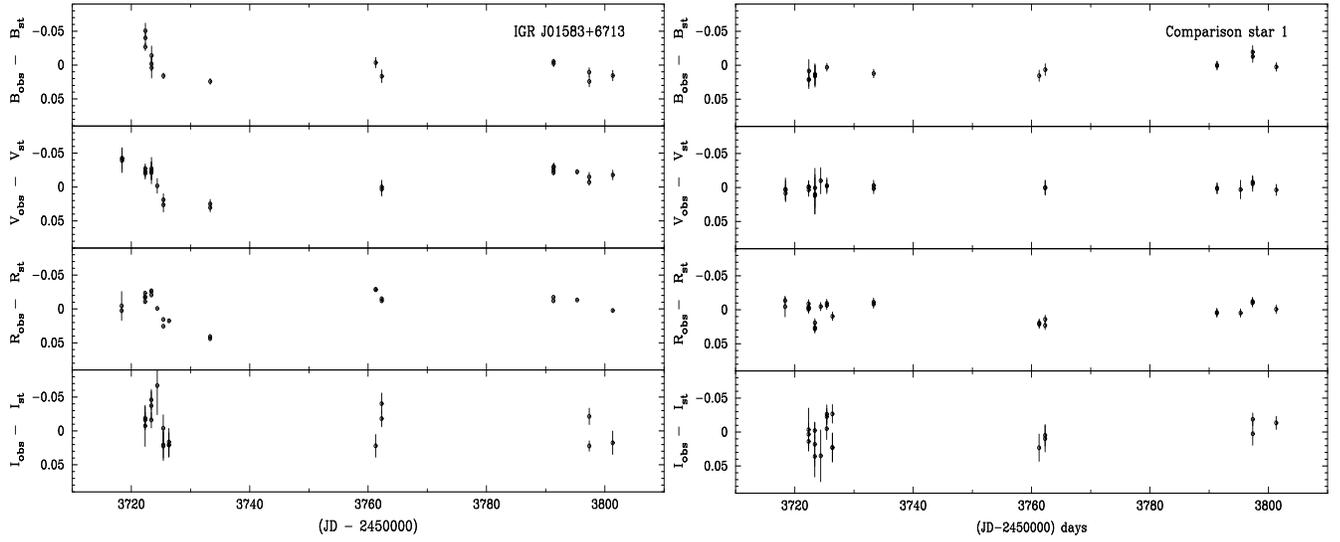}
\caption{The difference in the measured (B$_{obs}$, V$_{obs}$, R$_{obs}$ and I$_{obs}$) and 
standard (B$_{st}$, V$_{st}$, R$_{st}$ and  I$_{st}$) {\it BVRI} magnitudes of IGR J01583+6713 
and Comparison star 1 from JD 2453710 to JD 2453810.}
\end{figure}

\clearpage

\begin{figure}
\centering
\includegraphics[height=6.5in,angle =-90]{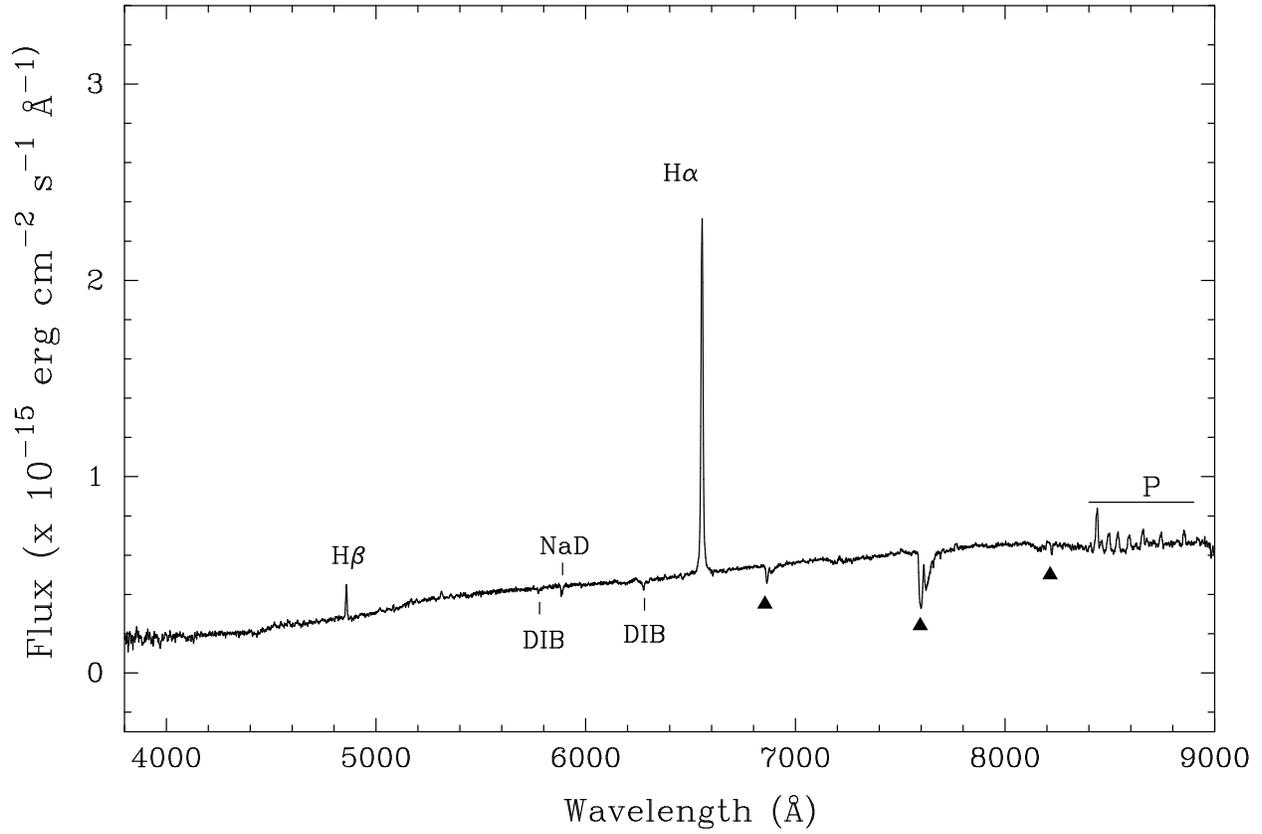}
\caption{Flux calibrated spectrum of IGR J01583+6713 taken on October 15, 2006. 
Grism 7 and Grism 8 spectrum are combined together to show it over a wavelength
range of 3800--9000 \AA. Diffuse Interstellar bands are marked as "DIB", 
telluric absorption bands are marked with a filled triangle and the Sodium doublet 
is marked as "NaD". "P" represents Paschen lines. H$\alpha$ and H$\beta$ are also marked.}
\end{figure}

\clearpage

\begin{figure}
\centering
\includegraphics[height=6in, angle =-90]{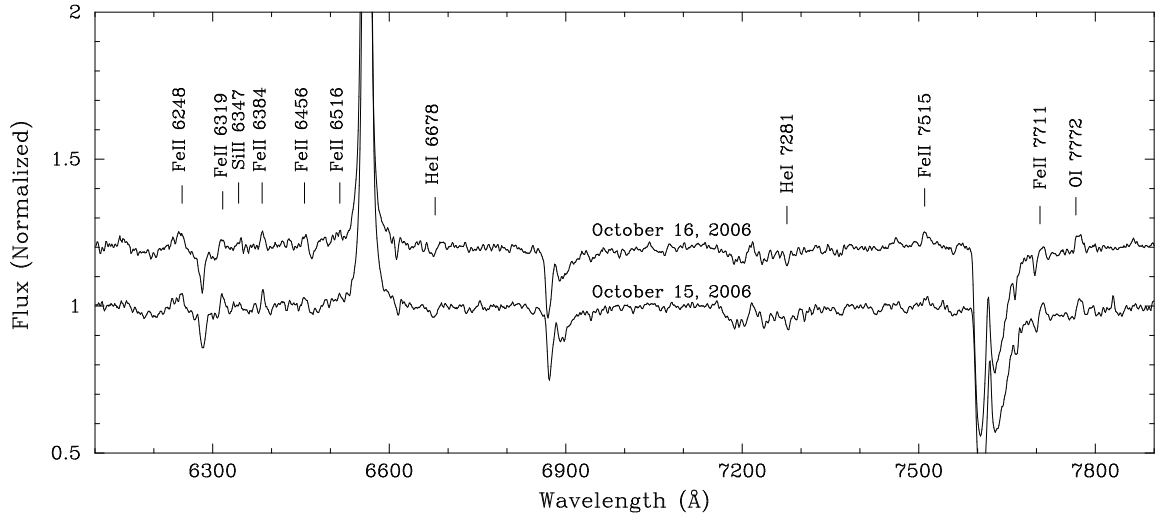}
\caption{Continuum spectrum of IGR J01583+6713 in the wavelength range of 6100--7900 \AA~taken on
October 15, 2006 and October 16, 2006. A few weakly identified features like FeII, HeI, SiII and OI 
are marked in the Figure.}
\end{figure}

\clearpage

\begin{figure}
\centering
\includegraphics[height=6.5in, angle =-90]{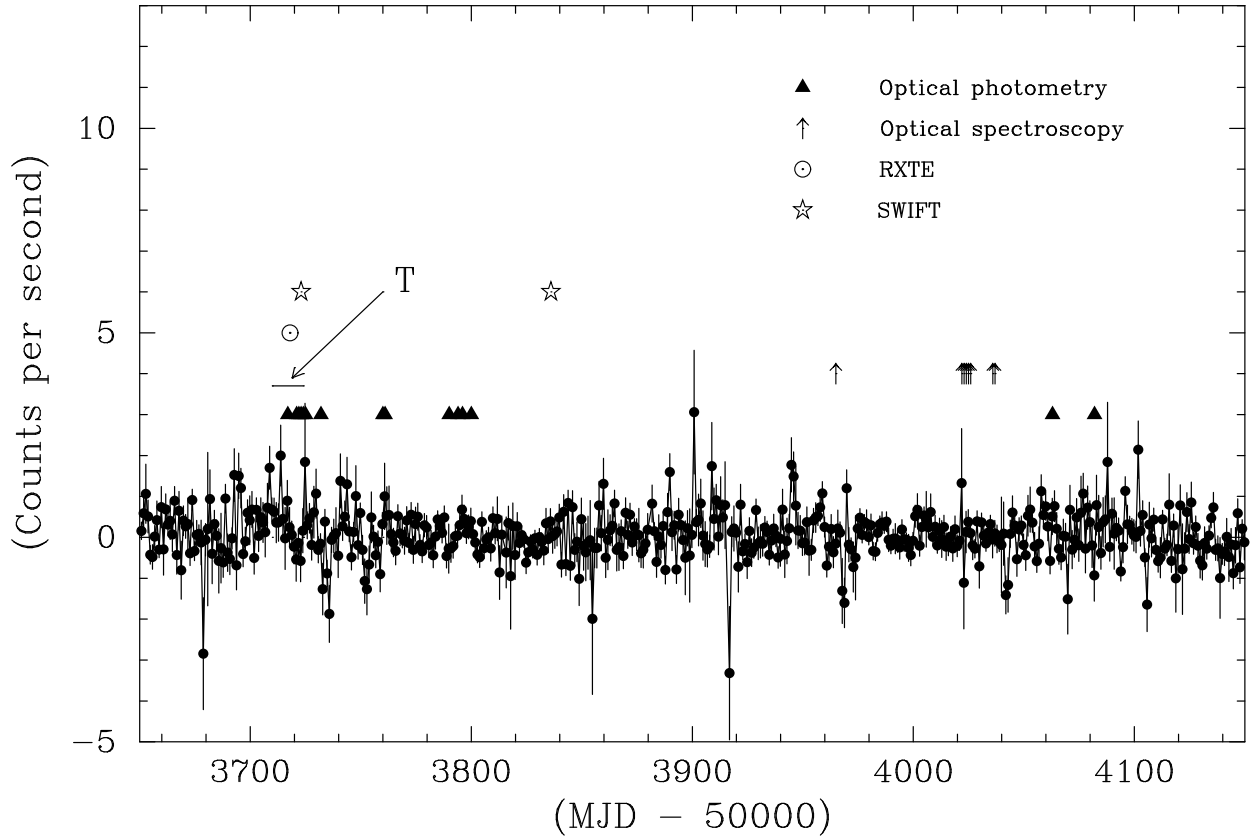}
\caption{ASM lightcurve of IGR J01583+6713 from MJD 53660 to MJD 54160
including the outburst observed by {\it  INTEGRAL} and {\it Swift} during MJD 
53710 to MJD 53720, marked as "T" in the figure. Also marked are optical 
photometric observations (by filled triangle), optical spectroscopic 
observations (by up-arrow), { \it RXTE} observations (by open circles 
with a dot inside) and {\it Swift} observations (by open star). }
\end{figure}

\clearpage

\begin{figure}
\centering
\includegraphics[height=6.5in]{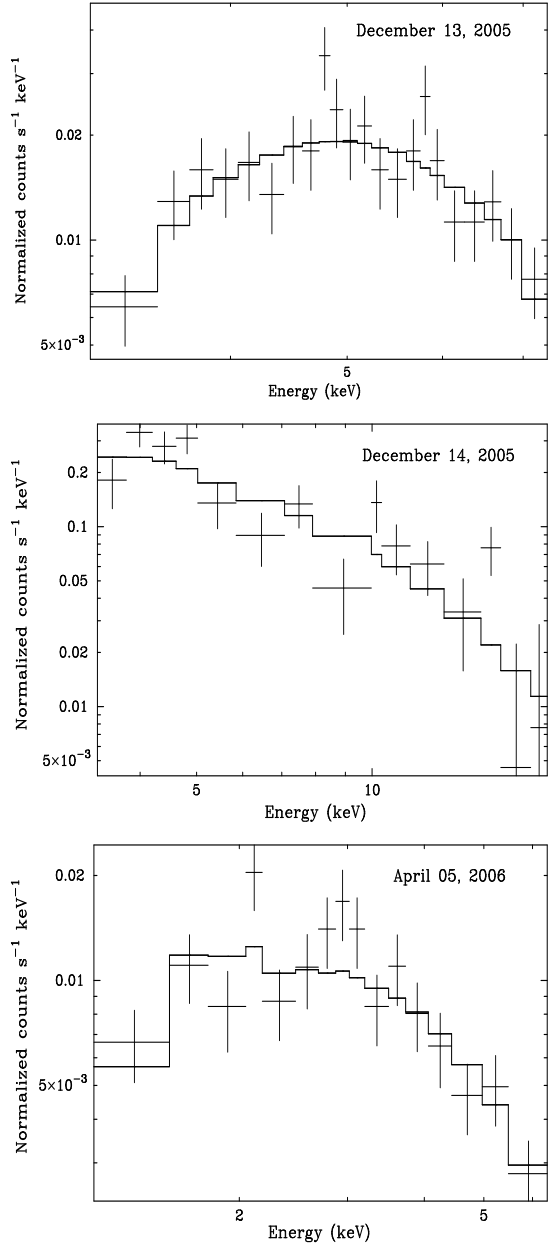}
\caption{X-ray spectrum of IGR J01583+6713. Top - { \it Swift} observations 
made on December 13, 2005, middle - {\it RXTE} observations on December 14, 2005, 
bottom -- {\it Swift} observations made on April 05, 2006. The points with error 
bars are the measured data points and the histograms are the respective best 
fitted model spectrum consisting of absorbed powerlaw model components, convolved 
with the respective telescope/detector responses.}
\end{figure}

\clearpage

\begin{figure}
\centering
\includegraphics[height=6.5in,angle =-90]{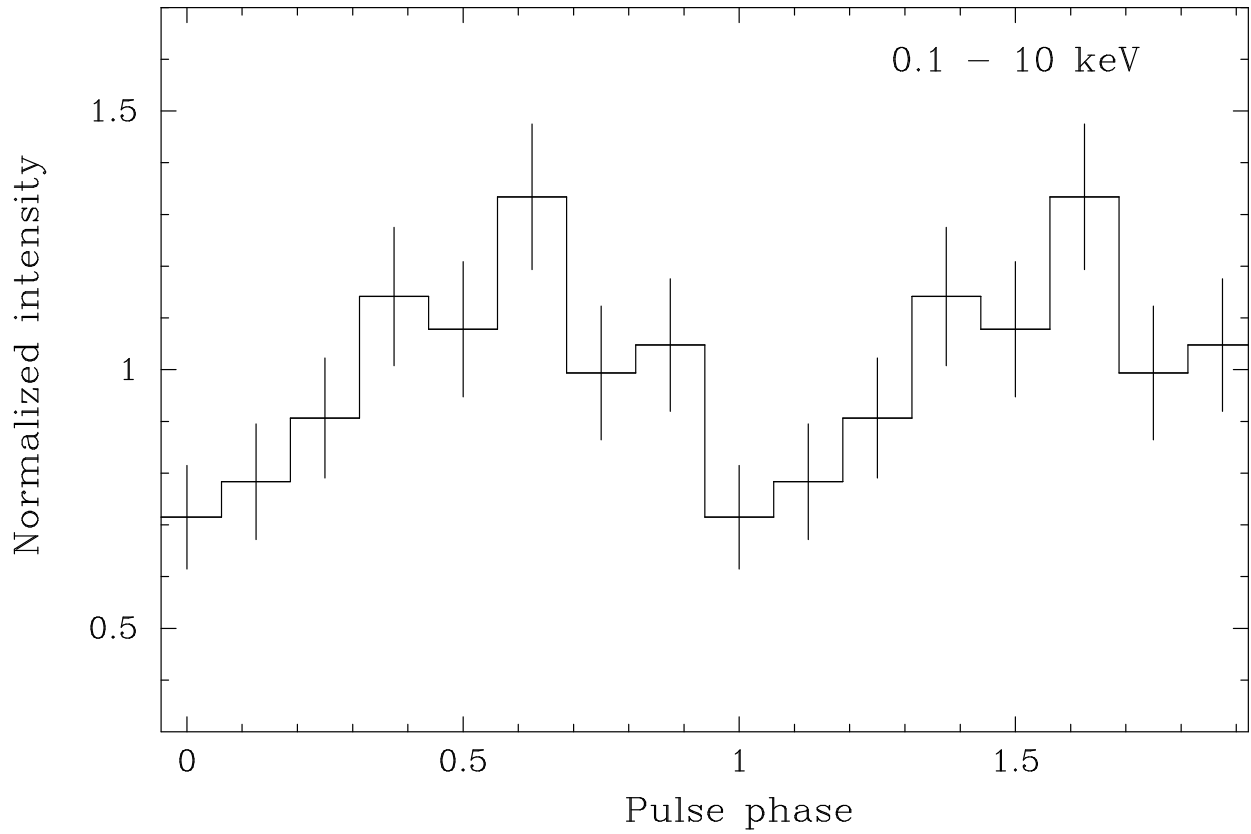}
\caption{Light curve of IGR J01583+6713 observed by  {\it Swift} on December 13, 2005, 
folded modulo 469.2 s.}
\end{figure}
\clearpage

\clearpage
\begin{figure}
\centering
\includegraphics[height=5in]{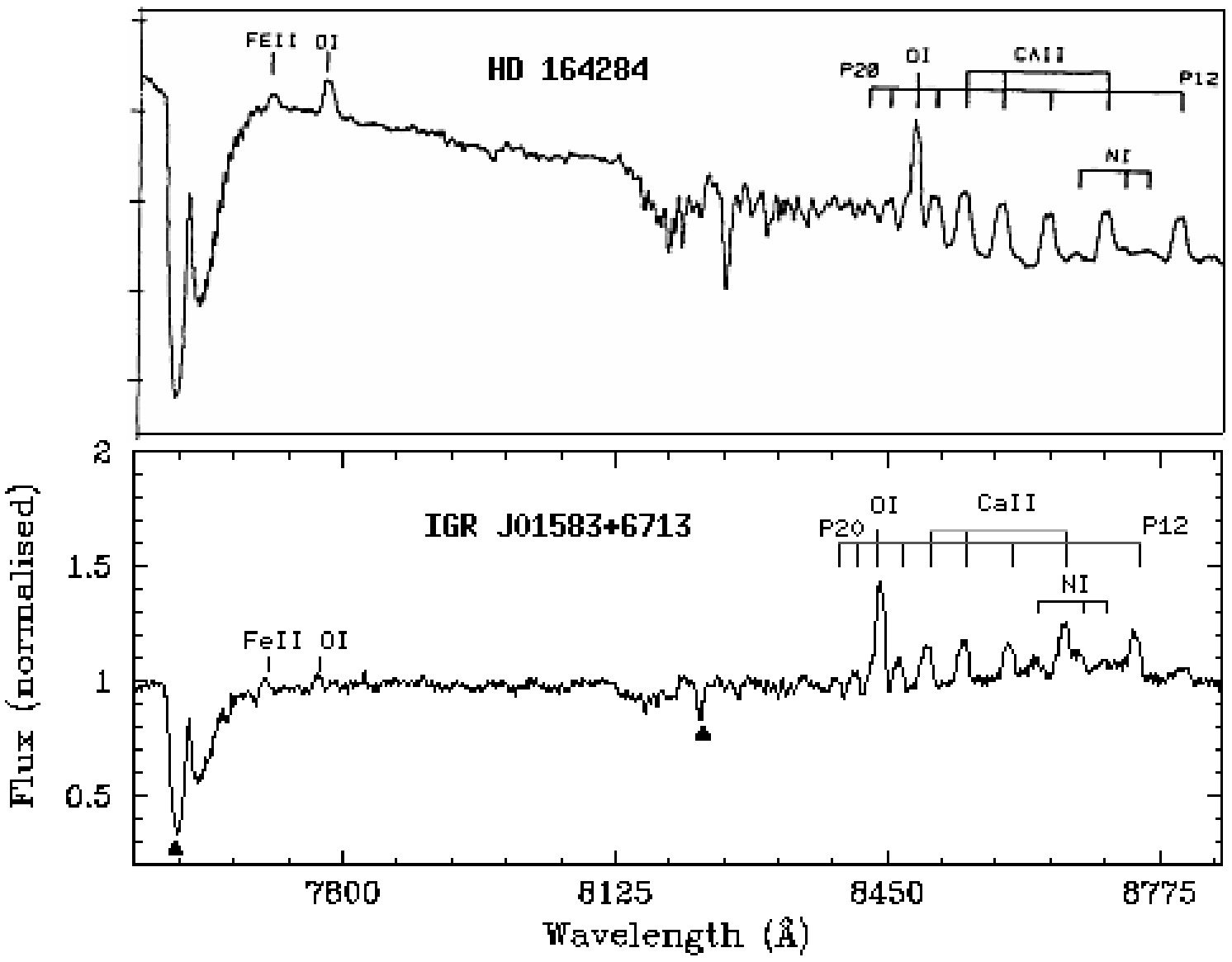}
\caption{Bottom - Flux Normalized spectrum of IGR J01583+6713, Top - Flux calibrated
spectrum of HD 164284 from Andrillat et al. (1988), a comparison. P12 to P20 are the 
Hydrogen Paschen lines from P12 to P20. FeII, OI, CaII and NI spectral features are 
also marked in the Figure.}
\end{figure}

\clearpage


\begin{thebibliography}{}
\bibitem{}Andrillat, Y., Jaschek, M., Jaschek, C., 1988, A\&AS, 72, 129
\bibitem{}Andrillat, A., Jaschek, M., Jaschek, C., 1990, A\&AS, 84, 11
\bibitem{}Bildsten, L., Chakrabarty, D. John, C. et al., 1997, ApJS, 113, 367
\bibitem{}Corbet, R. H. D., 1986, MNRAS, 220, 1047
\bibitem{}Fitzpatrick, E. L., 1999, PASP, 111, 63 
\bibitem{}Halpern, J. P., Tyagi, S., 2005, ATel, 681
\bibitem{}Horne, K., 1986, PASP, 98, 609
\bibitem{}Johnson, H. L., Morgan, W. W., 1954, ApJ, 119, 344
\bibitem{}Kennea, J. A., Racusin, J. L., Burrows, D. N., et al., 2005, ATel, 673
\bibitem{}Landolt, A. U., 1992, AJ, 104, 340 
\bibitem{}Leitherer, C., 1988, ApJ, 326, 356
\bibitem{}Lang, K. R. 1992, Astrophysical Data: Planets and Stars (New York: Springer-Verlag)
\bibitem{}Makishima, K. 1986, in The Physics of Accretion onto Compact Objects, ed. K. O. Mason, M. G. Watson, \& N. E. White (Berlin: Springer), 249
\bibitem{}Masetti, N., Pretorius, M. L., Palazzi, E., et al. 2006a, A\&A, 449, 1139
\bibitem{}Masetti, N., Bassani, L., Bazzano, A., et al. 2006b, A\&A, 455, 11
\bibitem{}Negueruela, I., Okazaki, A. T., Fabregat, J., Coe, M. J., Munari, U., Tomov, T., 2001, A\&A, 369, 117
\bibitem{}Negueruela, I., Schurch, M. P. E., A\&A, 2007, 461, 631
\bibitem{}Schmidt-Kaler, Th. 1982, Landolt-Bornstein: numerical data, ed. K. Schaifers, \& H. H. Voigt (Springer-Verlag)
\bibitem{}Steiner, C., Eckert, D., Mowlavi, N., et al. 2005, ATel, 672
\bibitem{}Stetson, P. B., 1987, PASP, 99, 191
\bibitem{}Stetson, P. B., 1992, JRASC, 86, 71 
\bibitem{}Wegner, W., 1994, MNRAS, 270, 229
\noindent
\end{thebibliography}
\end{document}